\documentclass[aps,pre,groupedaddress,amssymb,amsmath,pra,floatfix,showpacs]{revtex4}
\usepackage{epsfig}

\begin{document}
\title{Universal analytic properties of noise. Introducing the J-Matrix formalism}
\author{Daniel Bessis and Luca Perotti}
\affiliation{Department of Physics, Texas Southern University, Houston, Texas 77004 USA}
\date{\today}

\begin{abstract}
We propose a new method in the spectral analysis of noisy time-series data for damped oscillators.

From the Jacobi three terms recursive relation for the denominators of the Pad\'{e} Approximations built on the
well-known Z-transform of an infinite time-series, we build an Hilbert space operator, a J-Operator, where each
bound state (inside the unit circle in the complex plane) is simply associated to one damped oscillator while the
continuous spectrum of the J-Operator, which lies on the unit circle itself, is shown to represent the noise.

Signal and noise are thus clearly separated in the complex plane.

For a finite time series of length 2N, the J-operator is replaced by a finite order J-Matrix J$_{N}$, having N
eigenvalues which are time reversal covariant.

Different classes of input noise, such as blank (white and uniform), Gaussian and pink, are discussed in detail,
the J-Matrix formalism allowing us to efficiently calculate hundreds of poles of the Z-transform. {\it Evidence of
a universal behaviour in the final statistical distribution of the associated poles and zeros of the Z-transform is
shown. In particular the poles and zeros tend, when the length of the time series goes to infinity, to a uniform
angular distribution on the unit circle. Therefore at finite order, the roots of unity in the complex plane appear
to be noise attractors.}

We show that the Z-transform presents the exceptional feature of allowing {\it lossless undersampling} and how to
make use of this property.

A few basic examples are given to suggest the power of the proposed method.
\end{abstract}


\maketitle
\newpage

\section{Introduction}

Spectral analysis of highly noisy time-series data impacts such diverse fields as gravitational wave detection,
Nuclear Magnetic Resonance Spectroscopy as applied to nuclear waste, brain/breast cancer detection, oil detection
and other similar areas of application.

Experimental time-series are always affected by the presence of noise. As long as the signal to noise ratio is not
too poor, several filters are available to denoise the data within the framework of Fourier analysis and its
variants. All such techniques, on the other hand, have drawbacks.

Fourier analysis, being linear, treats noise on the same foot as the signal, and therefore is per se unable to
distinguish the two. In the case of stationary uncorrelated noise, Weiner-Khintchine theorem can be used to denoise
the data \cite{couch}, but it is unable to distinguish signal from correlated or non stationary noise.

Wavelets denoising methods require some knowledge of the signal to be found in order to avoid smoothing out the
signal itself \cite{donoho}.

Prony's method is tailored to analyze signals composed of damped oscillators, but it is known to be unstable in the
presence of noise. This is because -at least in its original formulation- it assumes an ``all-pole" system; in
other words, it is equivalent to the construction of a Pad\'{e} approximant (rational approximation) with constant
numerator. Having no zeros it is unable to model noise, as we shall see in section \ref{s3}. Modifications have
been proposed to stabilize it \cite{osborne}, but, apart from being cumbersome, no effort has been made to classify
the poles so as to distinguish the noise ones from the signal ones.

The matched filtering technique, which calculates cross-correlations between noisy detector outputs and reference
waveforms from a ``library", is only useful when the waveforms are well predicted theoretically \cite{Wainstein}.

Still, all the above techniques fail when the (average power) signal to noise ratio approaches $1$. It is exactly
this very high noise case we deal with here, presenting a denoising method based on the analytic properties of the
{\it Z-transform} (generating function) of the data \cite{zeta}.

The present paper is thus organized:

From a given infinite time series, we build -by way of Pad\'{e} Approximations of its Z-transform- a tridiagonal
Hilbert space operator, a J-Operator, which, for a finite time series of length 2N, is replaced by a finite order
J-Matrix J$_{N}$, having N  time reversal covariant eigenvalues. Such eigenvalues correspond to the poles of the
Z-transform itself.

We then introduce a specific class of signals made of an arbitrary number of damped oscillators. The poles of the
Z-transform of such signals are known to be {\it inside the unit circle} \cite{zeta}.

It is on the other hand known that a Taylor series with random coefficients has the unit circle as natural boundary
\cite{stei}.

For a time series of noisy damped oscillators, the spectrum of our J-Operator will therefore comprise a discrete
spectrum where each bound state (inside the unit circle) is simply associated to one damped oscillator; and a
continuous spectrum which represents the noise, and lies on the unit circle in the complex plane.

It is known that for a finite time series the natural boundary due to the noise is replaced by doublets of poles
and zeros \cite{dou1,dou2,dou3,dou4} surrounding the vicinity of the unit circle. We numerically study the
statistical distribution of such poles and zeros for {\it purely} noisy time-series, considering  different input
classes, such as blank (white and uniform), Gaussian and even coloured noise. We find compelling evidence of {\it
universal behaviour. In particular the poles and zeros tend, when the length of the time series goes to infinity,
to the roots of unity in the complex plane which appear to be noise attractors.}

As the Fourier Transform can be seen as the specialization of the Z-transform to the unit circle, the above result,
combined with the knowledge that the poles associated with damped oscillators are {\it inside} the unit cicle,
explains the poor behaviour of the Fourier Transform and the like while analyzing damped oscillators spectral data
corrupted by heavy noise.

We then prove that the Z-transform presents the exceptional feature of allowing {\it lossless undersampling}. This
allows us to use ``interlaced sampling" to obtain multiple time sequences from a single one and thus improve
detection through coincidence.

Finally, we give some simple examples to suggest the power of this approach. As is usually done in Nuclear Magnetic
Resonance Spectroscopy, we consider one or more damped oscillator signals repeatedly observed. By covering the
complex plane with a lattice and counting the ``non-zero paired" poles in each cell (that is poles which have no
associated zero in their vicinity), an approximate evaluation of both the detection probability and the false alarm
level for a single observation can be obtained. ``Interlaced sampling" can, on the other hand, significantly
improve performance even in the case of a single time sequence.

\section{Analytic properties of the Z-transform}

Given a time-series $s_{0},s_{1},s_{2},.....s_{n},.......$, it is a
standard practice \cite{zeta} to associate to it its {\it
Z-transform} defined as:

\begin{equation}
Z(z)=\sum_{n\geq 0}s_{n}z^{-n}  \label{1}
\end{equation}

Clearly, due to the boundedness of the time-series coefficients, this
function is at least analytic outside the unit disk.

\subsection{Construction of the tridiagonal Jacobi Matrix.}\label{s2}

We shall now introduce the Hilbert space operator whose resolvant matrix
element is precisely the Z-transform.

We are going to deal mainly with the $\left[ \frac{n-1}{n}\right] (z)$ Pad\'{e} Approximant to the $Z-transform$ of
a given time series \cite{pade} and write

\begin{equation}
\left[ \frac{n-1}{n}\right] (z)=\frac{N_{n-1}(z)}{D_{n}(z)}
\end{equation}

Where $D_{n}(z)$ is {\it a monic polynomial of degree} $n$ and $N_{n-1}(z)$
{\it a polynomial of degree} $n-1$ in $z.$

\qquad It is known \cite{baker} that $D_{n}(z)$ satisfy the Jacobi three terms
recursive relation

\begin{equation}
D_{k+1}(z)=\left[ z-A_{k}\right] D_{k}(z)-R_{k}D_{k-1}(z)\qquad
D_{-1}(z)=1\qquad D_{0}(z)=1\qquad
\end{equation}

where

\begin{equation}
A_{k}=-(a_{2k}+a_{2k+1})\qquad R_{k}=a_{2k-1}a_{2k}\qquad k\geq 1\qquad
a_{0}=0
\end{equation}

The $a_{k}$ being the coefficients of the Stieltjes continued fraction expansion of the $Z-transform$ .

This relation can be written as an eigenvalue problem by
introducing the tridiagonal $(N+1)\times (N+1)$ Jacobi Matrix

\begin{equation}
J_{N}=\begin{bmatrix} A_{0} & 1 & 0 & ... & 0 & 0 \\ R_{1} & A_{1} & 1 & 0 &
0 & 0 \\ 0 & R_{2} & A_{2} & 1 & ... & 0 \\ ... & & & & 1 & 0 \\ 0 & ... & 0
& R_{N-1} & A_{N-1} & 1 \\ 0 & 0 & ... & 0 & R_{N} & A_{N}\end{bmatrix}
\end{equation}

Then, when $D_{N+1}(z)=0$, the previous recursive relation can be written as

\begin{equation}
J_{N}V=zV
\end{equation}

where V is the column vector

\begin{equation}
V^{T}=\left[ D_{0}(z),D_{1}(z),...D_{N}(z),\right].
\end{equation}

{\it We thus see that the zeros of }$D_{N+1}(z)${\it \ are the eigenvalues (}$ z_{0},z_{1},...,z_{N}${\it ) of the
tridiagonal Jacobi Matrix of order N+1.} Therefore, the $Z-transform$ appears to be a matrix element of the
resolvant of the $J-Matrix$.

We also introduce the norm of the eigenvector:

\begin{equation}
\rho _{k}^{N}=%
\mathop{\displaystyle\sum}%
\limits_{p=0}^{p=N}\left\vert D_{p}(z_{k})\right\vert ^{2}
\end{equation}

\subsection{The case of infinite time-series}

\subsubsection{Damped oscillating signals Z-transform}

One of the most important application of the {\it Z-transform}
concerns damped oscillating signals, such as those found in nuclear
magnetic resonance data \cite{bel} and in burst gravitational waves
\cite{bur,ring}.

Let us start by considering the signal in the absence of noise: for
a finite ensemble of damped oscillators, the discretized data will
read

\begin{equation}
s_{k}=
\mathop{\displaystyle\sum}
\limits_{p}A_{p}e^{i\omega _{p}\frac{k}{N}T} \qquad
k=0,1,2,...,N-1\qquad \omega _{p}=2\pi f_{p}+i\alpha _{p}
\end{equation}
where $A_{p},f_{p},$ and $\alpha _{p}$ are the amplitude, the frequency and
the damping factor of the $p^{th}$ oscillator. $T$ is the recording time and
$N$ the number of data.

We now consider their {\it Z-transform} (\ref{1}) and allow both
the recording time $T$ and the number of data $N$ to go to
infinity, while keeping $T/N$ constant, so that (\ref{1}) becomes
the (formal) Taylor series :

\begin{equation}
Z(z)=
\mathop{\displaystyle\sum}
\limits_{k=0}^{+\infty }s_{k}z^{-k}  \label{tre}
\end{equation}

Combining the previous two equations, we get for the {\it
Z-transform} the expression:

{\large \bigskip }
\begin{equation}
Z(z)=\mathop{\displaystyle\sum}\limits_{k=0}^{+\infty }s_{k}z^{-k}=
\mathop{\displaystyle\sum}
\limits_{k=0}^{+\infty }z^{-k}\mathop{\displaystyle\sum}\limits_{p}A_{p}e^{i\omega _{p}\frac{k}{N}T}=
\mathop{\displaystyle\sum}
\limits_{p}A_{p}
\mathop{\displaystyle\sum}
\limits_{k=0}^{+\infty }(z^{-1}e^{i\omega _{p}\frac{T}{N}})^{k}=\mathop{\displaystyle\sum}
\limits_{p}\frac{A_{p}}{1-z^{-1}e^{i\omega _{p}\frac{T}{N}}}
\end{equation}

From this we immediately see that, in this noiseless case, the {\it
Z-transform} is a rational fraction in $z$, the poles of which are

\begin{equation}
z_{p}=e^{i\omega _{p}\frac{T}{N}}
\end{equation}

Notice that all poles have to lie strictly inside the unit circle because $%
\mathop{\rm Im} \omega_{p}>0$. The residues are

\begin{equation}
\rho_{p} =z_{p}A_{p}
\end{equation}

Therefore, after identifying the poles $z_{p}$ and residues $\rho
_{p}$ of the {\it Z-transform}, we get the quantities of interest,
which are the frequencies

\begin{equation}
f_{p}=\frac{N}{T}\frac{\arg z_{p}}{2\pi }\label{freq}
\end{equation}%
the damping factors
\begin{equation}
\alpha_{p}=\frac{N}{T}\log{|z_{p}|}
\end{equation}%
and the amplitudes
\begin{equation}
A_{p}=\frac{\rho _{p}}{z_{p}}\label{ampl}
\end{equation}%
of the damped oscillators of the signal.

This straightforward calculation shows that for a sum of
oscillating damped signals, the Z-transform associated to its
time-series is a sum of poles in the complex plane. The position of
each pole is simply linked to the damping factor and the frequency
of each of the oscillators. Also, it is important to note that all
these poles lie {\it strictly} inside the unit disk.

\subsubsection{Noise Z-transform}

Due to the linearity of the {\it Z-transform}, the {\it Z-transform} of a
noisy time-series will be the sum of the {\it Z-transform} of the signal
plus the {\it Z-transform} of the noise.

{\it It is therefore of the uttermost importance to get a deep insight
inside the analytic structure of a purely noisy time-series.}

The following theorem is the key to the answer to this question.

{\bf Steinhaus theorem:} A Taylor series with random coefficients
has, {\it with probability one (that is: except for a set of
measure zero),} the unit circle as natural boundary \cite{stei}.

\subsubsection{The case of noisy data}\label{s3}

Combining the previous statements, we get the following simple but hitherto unrecognized fundamental result.

The {\it Z-transform} of a finite number of damped oscillations is,
in the presence of noise, an analytic function outside the unit
disk, plus a rational function having a finite number of poles
strictly inside of the unit disk, each of them representing one
damped oscillation.

We can state the following fundamental theorem.

{\bf Theorem }(Bessis-Perotti).

To any noisy time series can be associated a tridiagonal Hilbert space operator, its $J-Operator$, extension when
$N\rightarrow \infty$ of the previously defined $J_{N}$. With probability one (that is: except for a set of input
data series of measure zero), the spectrum of this operator is made of two parts:

An essential spectrum with support the unit circle. This spectrum
is associated with the noise (uncorrelated part of the signal). The
corresponding eigenfunctions have {\it infinite norm}. At a finite
order, when the infinite $J-Matrix$ is truncated, it decomposes
into the Froissart poles of the Froissart doublets which will be described in the following section.

A discrete spectrum, made of a finite number of poles inside the
unit circle: each pole represents a component of the signal made of
a finite number of damped oscillators. The corresponding
eigenfunctions have {\it finite norm}.

\subsection{The case of finite time-series}

Up to now we have been dealing with time-series of infinite length, we come
now to the more realistic case where the time-series has a finite length $N$.

The problem of constructing the {\it Z-transform} $Z(z)$ of such
finite time-series is a standard problem in Mathematics known as
the Pad\'{e} Approximation construction \cite{pade}.

\subsubsection{The special case of non noisy finite time-series}

If P is the number of damped oscillators, as soon the length of the
time-series is $N \geq 2P+1$, the diagonal Pad\'{e} approximant
$P/P$ gives an elegant solution to the above problem.

\subsubsection{The general case}

In this case, M. Froissart \cite{fro} has shown that the natural boundary generated by the noise is approximated by
doublets of poles and zeros (Froissart doublets - see e. g. \cite{dou1,dou2,dou3,dou4} ) surrounding the vicinity
of the unit circle. The mean distance of the members of each doublet being of the order of the noise magnitude and
{\it we conjecture that it decreases exponentially with the length of the signal.}

\section{Universality properties of the noise part of the signal}

{\it Here, we are mainly interested in studying the statistical distribution
(both radial and angular) of the Froissart doublets as a function of the
input noise statistical properties and the length N of the time-series.}

The most {\bf unexpected} result is the universality of the
distribution: the radial distribution in the purely noisy case is
universally Lorentzian, the phase distribution is uniform
(approaching the $[N/2]$ roots of unity when $N$ goes to infinity).

Due to this property of {\it universality,} the only parameter left is the
width of the Lorentzian, which depends on the length $N$ of the time-series.
A conjecture, based on heuristic arguments, about the functional dependence
of this width on $N$ will be provided in the sequel.

\subsection{Experimental evidence for universality}

We calculated the radial distribution of the diagonal and sub
diagonal Pad\'{e} approximants for time-series of up to $600$ data,
which allow the construction of Pad\'{e} approximants up to
$300/300$.

To test universality, we have been studying $7$ different kinds of noise:

1) the complex input data have uniform distribution both in modulus and
phase.

2) the complex input data have uniform distribution inside a square domain.

3) the complex input data have uniform distribution inside a circle domain.

4) standard pink noise.

5) the complex input data have Gaussian distributions.

6) the complex input data are correlated following an autoregressive moving
average (ARMA) model.

7) the input data are real and have a Gaussian distribution.

\begin{figure}[htbp]
\centering\epsfig{file=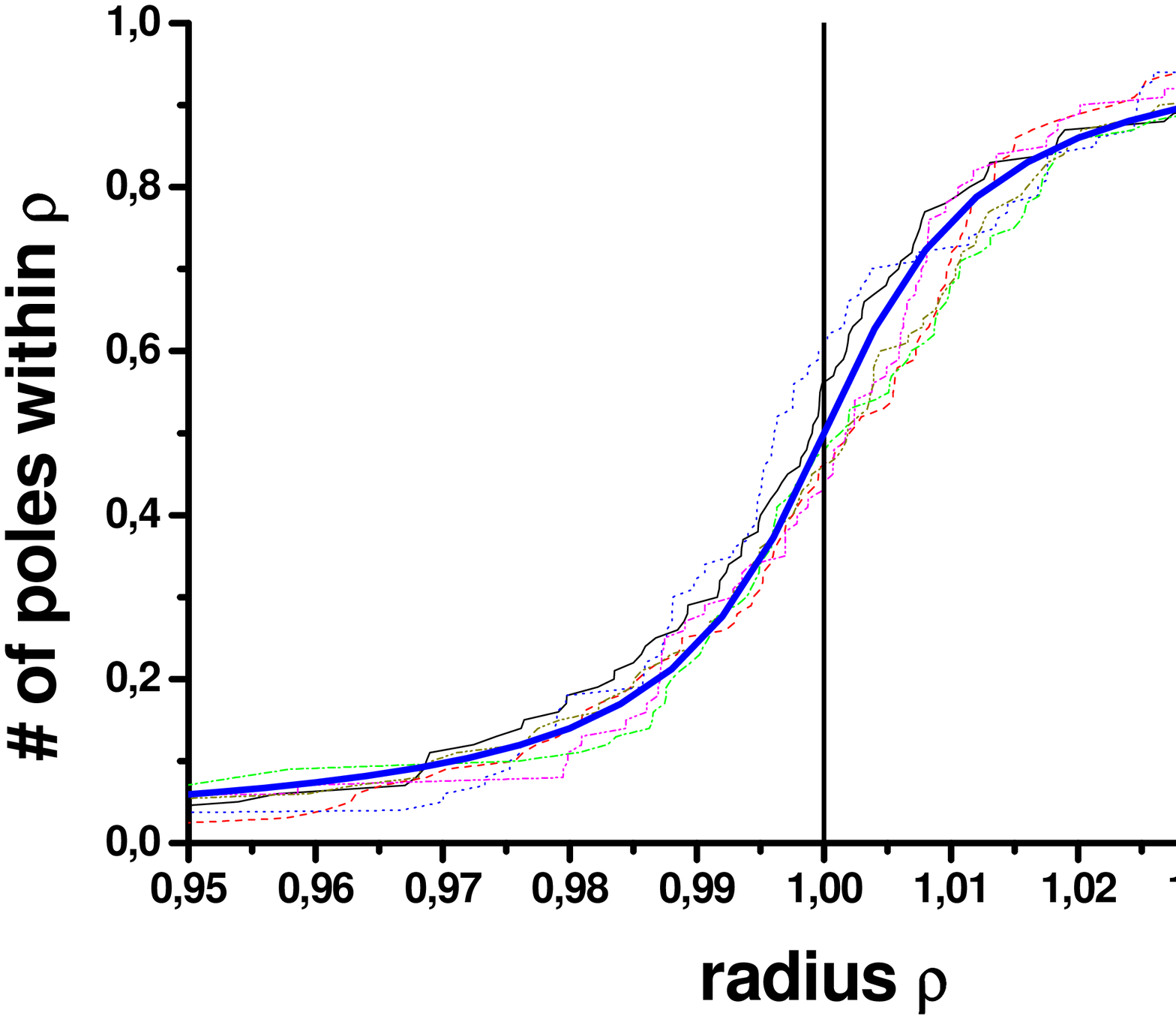,width=0.7\linewidth}
\vspace{-0.6in}
\caption{Radial cumulative distribution of the Froissart
poles of the [100/100] Pad\'{e} approximant for different kinds of
noise (see text for a full description). Full line, black: type 1;
dash, red: type 2; dot, blue: type 3; dash-dot, green: type 4;
dash-dot-dot, magenta and dark yellow: two different cases of type
5. The thick full line is a fitting by a Lorentzian cumulative
distribution ($(1/ \pi) \arctan[(\rho-1)/W]+1/2$, where $W$ is the
width of he distribution).}
\label{fig1a}
\end{figure}

Typical results for the {\it single run} radial distribution of the Pad\'{e} approximant poles in some of the above
seven cases are shown in Figure (\ref{fig1a}): {\it in all cases the behaviour is Lorentzian and centered on 1.}

\begin{figure}[htbp]
\centering\epsfig{file=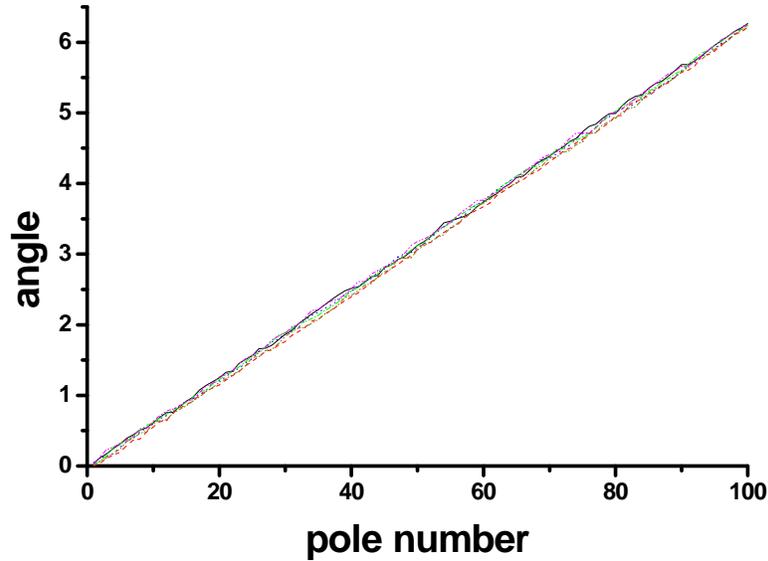,width=0.7\linewidth}
\vspace{-0.6in}
\caption{Phase distribution of the Froissart poles of the
[100/100] Pad\'{e} approximant for different kinds of noise (see
text for a full description). Full line, black: type 1; dash, red:
type 2; dot, blue: type 3; dash-dot, green: type 4; dash-dot-dot,
magenta and dark yellow: two different cases of type 5.}
\label{fig1b}
\end{figure}

Figure (\ref{fig1b}) presents instead the phase distributions of the same
Froissart poles: {\it in all cases the poles are uniformly distributed
between $0$ and $2\pi $,} {\it the deviation being Gaussian}.

The graphs shown refer to time-series of $200$ data: the case for which we
have run the highest number of tests, varying several times the seeds of the
random number generators used for each of the $7$ kinds of noise.

Analogous results apply to the zeros of the Pad\'{e} approximants, with the difference that the radial distribution
is centered {\it on a point larger than 1} which approaches unity when $N$ goes to infinity (at present, we have no
explanation for this slightly unsymmetrical behaviour between the poles and zeros statistical distributions).

We have recently gained evidence that Universality extends to the case of noise statistical distributions having
infinite second moment and, even, infinite mean. We shall report the details in a forthcoming paper.

For the very special case of Gaussian uniform noise, a mathematical proof of our results can be found in Ref.
\cite{barone}.

\subsection{The width of the universal radial Lorentz distribution}

We are left with the problem of finding an expression for the width of the
universal radial distribution of both the poles and zeros and for the
distance of the center of the radial distribution of the zeros as a function
of the time-series length N.

On heuristic arguments, we propose for all three cases the following formula
\begin{equation}
W\simeq \alpha {\frac{{\ln {(N)}}}{{N}}}  \label{2}
\end{equation}

\begin{figure}[htbp]
\centering\epsfig{file=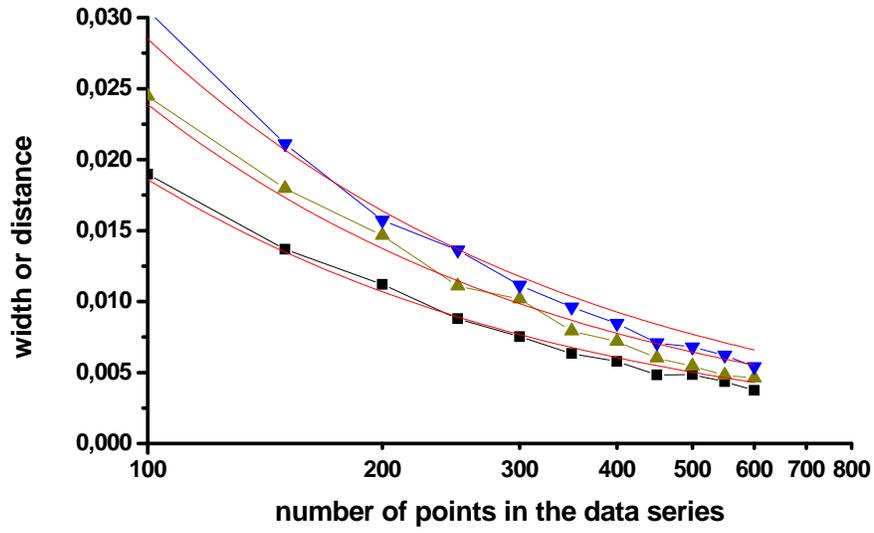,width=0.7\linewidth}
\vspace{-0.6in}
\caption{The width
of the universal radial distribution of both the poles (squares, black) and zeros (up triangles, dark yellow), and
the distance from unity of the center of the radial distribution of the zeros(down triangles, blue) as a function
of the time-series length N. Also shown is their fitting to \ref{2}. The global multiplicative factor $\alpha$ is,
respectively, $0.404 \pm 0.005$, $0.52 \pm 0.01$, and $0.62 \pm 0.01$.}
\label{fig2}
\end{figure}

Figure (\ref{fig2}) shows a fit to the above formula of said three
quantities. The agreement appears quite good and seems to improve when
increasing $N$.

\section{First conclusions}

Spectral analysis has been dominated by the Fourier transform (and its
variants), which is the ideal tool when noise is not too important.

The major drawback of the Fourier transform is its {\it linearity} which
prevents it to distinguish noise from signal because the Fourier Transform
treats the noise on the same foot as the signal, and therefore leaves noise
intact.

{\it Furthermore it is clear that the discrete Fourier Transform is nothing
but the specialization of the Z-transform to specific points of the complex
plane which are the roots of unity.} This is dramatic because, as we have
shown in our presentation, {\it the roots of unity are the noise attractors
of the complex plane.}

{\it This explains the poor results of data spectral analysis in the
presence of heavy noise produced by the FFT.}

We therefore suggest to substitute the {\it Z-transform} for the discrete Fourier Transform when analyzing highly
corrupted time-series data.

\section{Beyond the Fourier Analysis: methods for extracting Spectra from Noisy Data.}\label{s4}

In order to make a clear distinction among the poles of the Z-transform of a noisy time series of length $2N+2$,
and identify the eigenvalues corresponding to the discrete spectrum, we can make use of several properties of the
Z-transform. Each method of those we are going to present addressing a different property, a comparison of the
results of all methods applicable to any specific time-series is likely to be the optimal choce.

For some of the methods delineated below, we need to find both poles and zeros of the Z-transform. For the former
ones, we compute the $(N+1)\times (N+1)$ Jacobi Matrix associated to the denominator of the Z-transform we
described in section \ref{s2} and then diagonalize it. The latter ones can be calculated diagonalizing the $N
\times N$ Jacobi Matrix for the numerator of the Z- transform Pad\'{e} Approximant, which is simply the denominator
Jacobi Matrix with the first column and first row removed. The procedure for computing both zeros and poles of the
Z-transform therefore consists in the diagonalization of an already tridiagonal matrix; this makes it
computationally extremely efficient, and allows us to easily calculate hundeds of poles and zeros.

\subsection{Cleaning of Froissart doublets}

A first denoising method consists in removing from the complex plane the poles which can be identified as part of
Froissart doublets, due to their proximity to zeros of the Z-transform.

This can be done by ordering the poles and zeros in couples in order of increasing distance between the two, and
keeping only those whose distance is higher than a given value $\epsilon$ (usually $\epsilon \ge 0.2$).

\subsection{Variational principle}

We can on the other hand, making use of the results of section \ref{s2}, search the eigenvalues of Jacobi Matrix
associated to the denominator of the Z-transform having {\it modulus strictly smaller than one} and for which the
norm of the eigenvector have local minima. Such eigenvalues correspond to the frequencies of the signal.

\subsection{Stationariety}

It is also possible to compare Pad\'{e} approximants of different order and look for stable ``non-zero paired"
poles inside the unit circle: these will be the signal poles, while the non stationary poles will be linked to the
noise. This method is particularly useful in dealing with poles in the tails of the noise pole distribution.

\subsection{Undersampling}

Another method makes use of an exceptional feature, never reported up to now, of Pad\'{e} Approximations with
respect to undersampling.

Emmanuel Candes \cite{candes} has extended the Nyquist-Shannon sampling theorem \cite{shanon} and shown that it was
possible under some circumstances to use bandwidth smaller than the ones requested by the Nyquist theorem
(sufficient condition only but not necessary) without losing information.

It is therefore interesting to look for classes of signals for which the corresponding Z-transforms allow
undersampling {\it without any loss of information. We are going to show that the class of damped oscillators
allows lossless undersampling.}

Suppose that in (\ref{tre}), we undersample the time series by taking only the $s_{k}$ with $k=mn$ where $m$ is
given fixed, while $n$ runs from zero to infinity to start with. We get the undersampled $ Z_{m}-transform$

\begin{equation}
Z_{m}(z)=
\mathop{\displaystyle\sum}%
\limits_{n=0}^{+\infty }s_{mn}z^{-n}
\end{equation}
or
\begin{equation}
Z_{m}(z)=%
\mathop{\displaystyle\sum}%
\limits_{n=0}^{+\infty }s_{mn}z^{-n}=%
\mathop{\displaystyle\sum}%
\limits_{n=0}^{+\infty }z^{-n}%
\mathop{\displaystyle\sum}%
\limits_{p}A_{p}e^{i\omega _{p}\frac{mn}{N}T}=%
\mathop{\displaystyle\sum}%
\limits_{p}A_{p}%
\mathop{\displaystyle\sum}%
\limits_{n=0}^{+\infty }(z^{-1}e^{im\omega _{p}\frac{T}{N}})^{k}=%
\mathop{\displaystyle\sum}%
\limits_{p}\frac{A_{p}}{1-z^{-1}e^{im\omega _{p}\frac{T}{N}}}
\end{equation}

We see that the Pad\'{e} Approximation $P/P$ again gives the exact solution to the undersampled time series. So
provided we pick $N=2P+1$ equispatiated terms in the time series, we are able to rebuild it exactly.

We point out here that we assume to be dealing with data series such that the periods of the signal oscillations to
be reconstructed cover several data points. This means that, as long as $m$ is less than the number of data points
per period of each of the signal oscillators, the reconstructed signal frequencies will be given by the $m$th roots
of the poles $z_p$ having the lowest phases.

The above result suggests another technique -which we call ``interlaced sampling"- to improve sensitivity when only
a single data time-sequence is available. Taking advantage of the undersampling properties of the $Z-transform$, we
divide the data in $m$ undersampled sequences, the first one comprising the points $1,m+1,2m+1,3m+1,...$, the
second one comprising the points $2,m+2,2m+2,3m+2,...$ and and so on. To first oder, the ratio of the width of the
pole distribution around the unit circle to the distance of the signal pole from it does not change, but instead of
a single signal pole, we do now get a cluster of $m$ poles, which are much easier to detect. Preliminary tests have
given encouraging results.

\section{Application of the method to explicit examples}

We shall now proceed to describe some simple examples. The actual procedure we use to reconstruct the signal is
very basic, but sufficient for our present aim of showing the potentiality of our proposed system. Plain Fortran
was used to write the code implementing the procedure described in the following.

\subsection{A first example}

As a first example, we consider a damped signal at zero frequency. The chosen damping factor reduces the signal by
a factor $10^{-6}$ over $300$ signal points, resulting in a pole at $(0.95238,0.0)$. Noise is of the type 1) above
(the complex input data have uniform distribution both in modulus and phase).

A signal series of $300$ points is relatively short. Since the separation of
signal and noise is only univocal in the limit of an infinite series, and
-moreover- a single pole off the unit circle could be due to the particular
noise sequence used, we repeat the sequence several times, as is for example usually done in Nuclear Magnetic
Resonance Spectroscopy. In the present case, we repeat the sequence $40$ times, each time changing the seeds for
our noise generator, and search for a {\it stable} pole inside the unit circle with no zero close to it.

For each of the $40$ data sequences we clean most of the Froissart doublets according the procedure delineated
above: we calculate the distance of each pole from all the zeros, arrange the pole-zeros couples in order of
increasing distance and keep only the uncoupled pole and the poles whose distance from the coupled zeros is higher
than a set value. We find that cut distances between $0.6$ and $0.8$ -which leaves us with two or three ``non-zero
paired" poles per run- give good results for the range of signal/noise ratios we explored; for higher signal/noise
ratios good results can be obtained with much smaller cut distances.

Directly plot of these poles (and their coupled zeros) for all $40$ runs, can be misleading to the eye. To better
see areas of concentration of poles, we divide the complex plane in a number of square boxes and plot in three
dimensions the number of poles minus the number of coupled zeros in each box. For the tests we present, the boxes
were of side $0.1$.

\begin{figure}[htbp]
\centering\epsfig{file=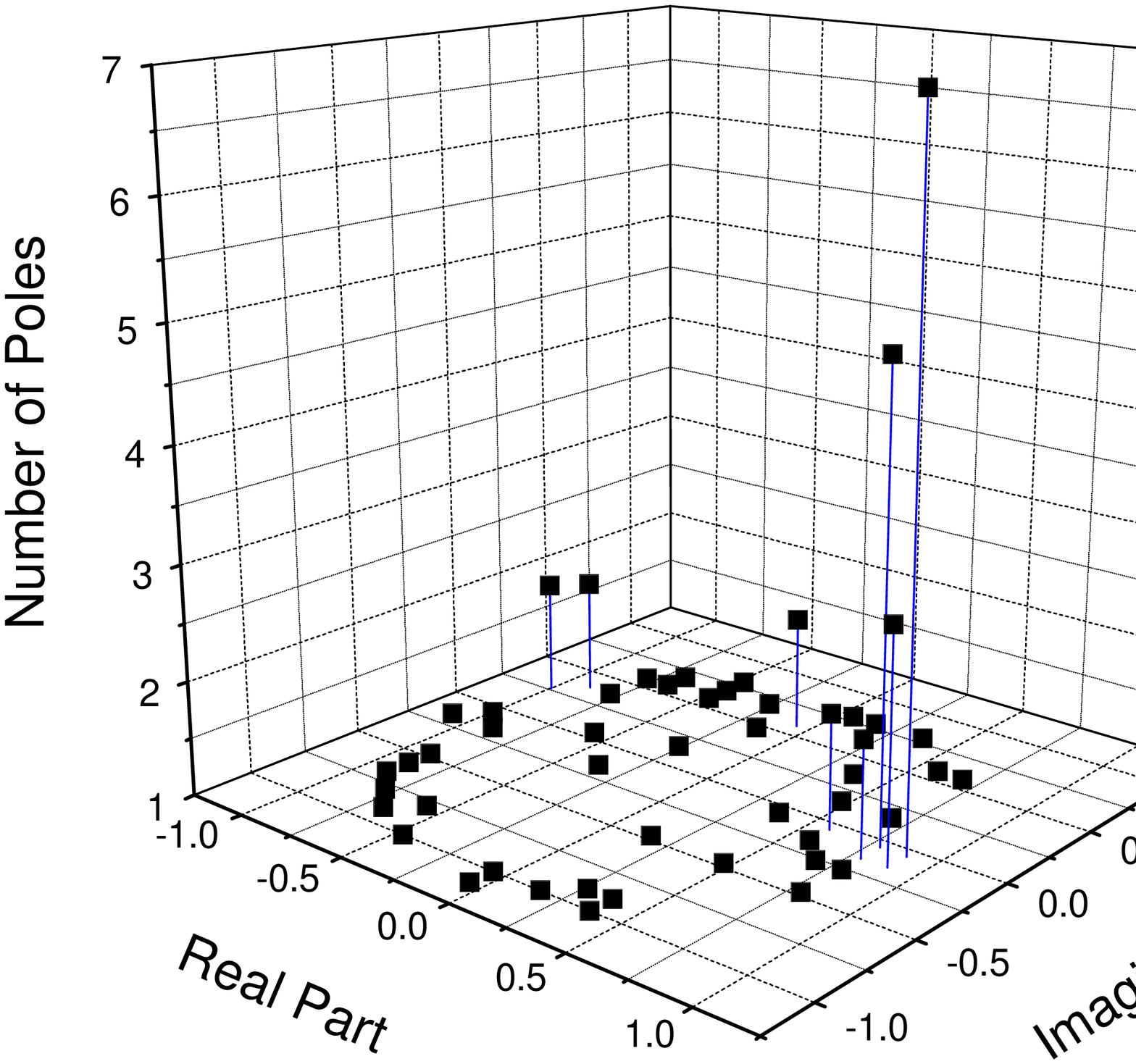,width=0.7\linewidth}
\vspace{-0.9in}
\caption{Signal/noise ratio of approximately $-10db$. Distribution of
the [149/150] Pad\'{e} approximant poles distant from their coupled
zeros more than $0.75$. The vertical axis represents the number of
poles in each box of side $0.1$. The highest peak corresponds to
the signal pole at $z=(0.95238,0.0)$.}
\label{fig3a}
\end{figure}

\begin{figure}[htbp]
\centering\epsfig{file=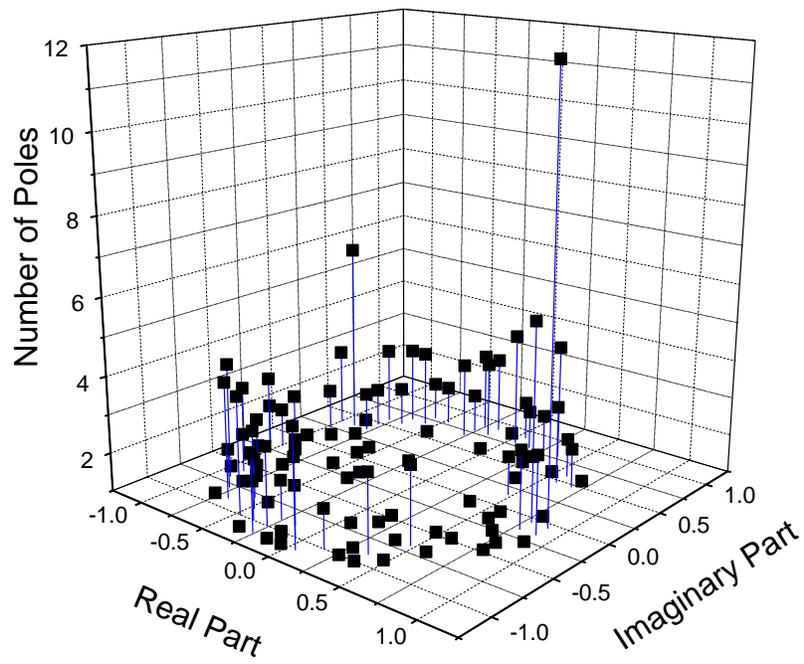,width=0.7\linewidth}
\vspace{-0.6in}
\caption{Same as Figure \ref{fig3a} above but the minimum pole-zero distance has been reduced to $0.2$.}
\label{fig3b}
\end{figure}

Figures (\ref{fig3a}) and (\ref{fig3b}) show the case of an average power signal/noise ratio (SNR) of approximately
$-10db$ (matched filtering SNR equal to $5.5$), using a cut distance of
-respectively- $0.75$ and $0.2$: in both cases the highest peak is at the
position of the signal pole.  Note that this can be considered a worst case scenario: the signal pole lies close to
a ``four corners point". The reconstructed poles therefore are distributed over four boxes: for a cut distance of
$0.75$ the four peaks sum to $17$, while the height of the highest ``fake" peak is only $2$.For a cut distance of
$0.2$ the four peaks sum to $22$, but the height of the highest ``fake" peak is now $6$. This suggests that
reducing the cut distance this much is not convenient: going from $0.75$ to $0.2$ improves the detection
probability from about $40\%$ to about $50\%$, but at the same time forces us to increase the false alarm threshold
by a factor $3$ (from $2$ to $6$ out of $40$). For the examples that follow we shall therefore show only the
results obtained using $0.75$ as our cut distance.

\begin{figure}[htbp]
\centering\epsfig{file=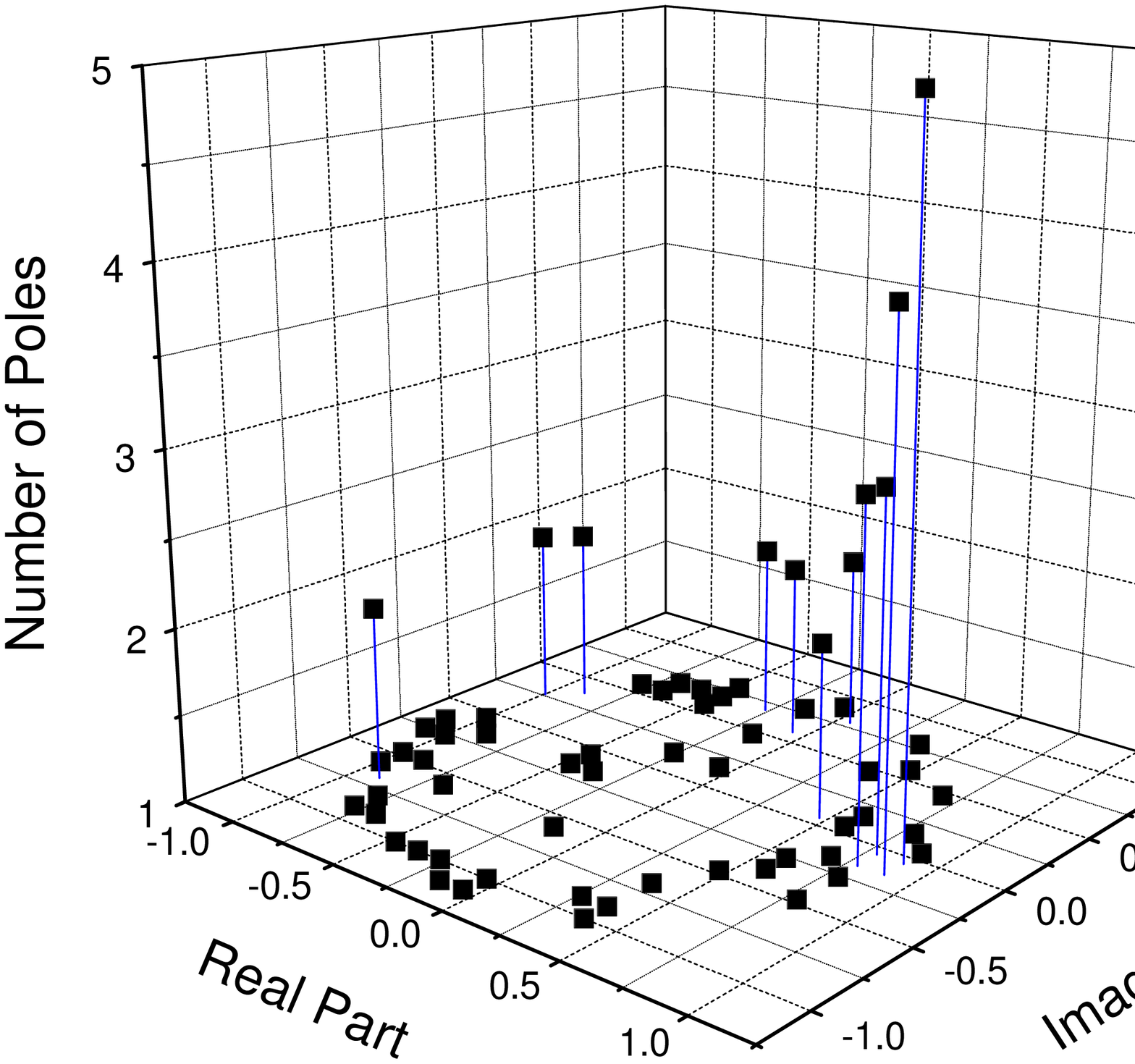,width=0.7\linewidth}
\vspace{-0.9in}
\caption{Signal/noise ratio of approximately $-15db$. Distribution of
the [149/150] Pad\'{e} approximant poles distant from their coupled
zeros more than $0.75$. The vertical axis represents the number of
poles in each box of side $0.1$. The highest peak corresponds to
the signal pole at $z=(0.95238,0.0)$.}
\label{fig4}
\end{figure}

Reducing the average signal/noise ratio to approximately $-15db$ (matched filtering SNR equal to $3.0$), the peak
is still the highest, but ``fake" peaks far from it become more visible, as can be seen in Figure (\ref{fig4}).
This is mostly due to a higher noise induced spread of the reconstructed poles: if we sum the four signal peaks we
get $15$, which is not much less than the $17$ of the previous case (detection probability is still around $40\%$),
especially considering that the height of the highest ``fake" peak is still $2$ which means we can keep the same
false alarm threshold.

\begin{figure}[htbp]
\centering\epsfig{file=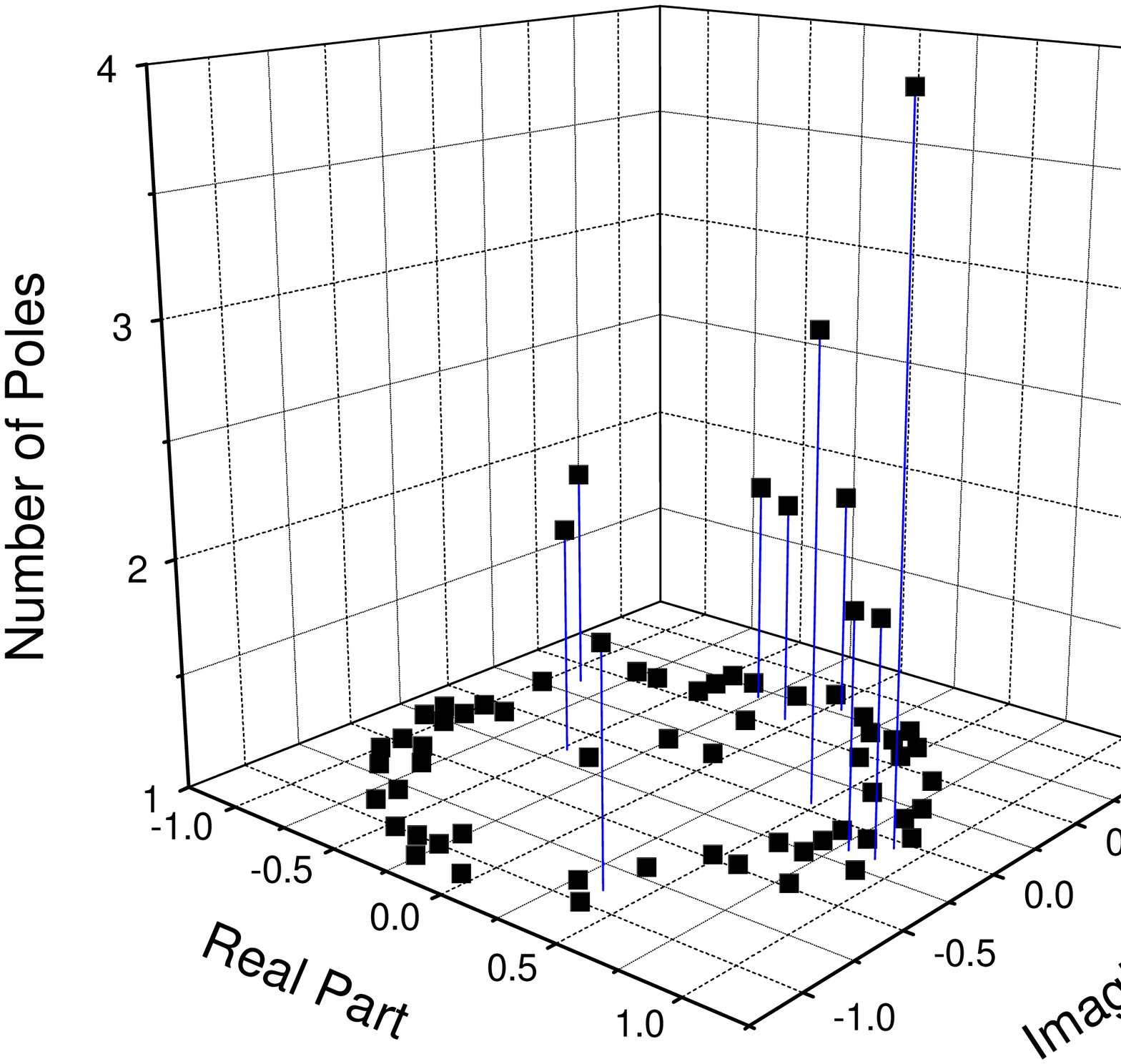,width=0.7\linewidth}
\vspace{-0.6in}
\caption{Signal/noise ratio of approximately $-18db$. Distribution of
the [149/150] Pad\'{e} approximant poles distant from their coupled
zeros more than $0.75$. The vertical axis represents the number of
poles in each box of side $0.1$. The highest peak corresponds to
the signal pole at $z=(0.95238,0.0)$.}
\label{fig5}
\end{figure}

Figure (\ref{fig5}) finally shows the case of an average signal/noise ratio of approximately $-18db$ (matched
filtering SNR equal to $2.2$): the peak at the signal pole is still the highest, but peaks far from it (the height
of the highest still being $2$)are now substantial compared to it and to the sum of the four signal peaks which is
now only $9$. The false alarm theshold does not need to be changed, but detection probability is now down to
$25\%$.

\subsection{A more complicated signal}

\begin{figure}[htbp]
\centering\epsfig{file=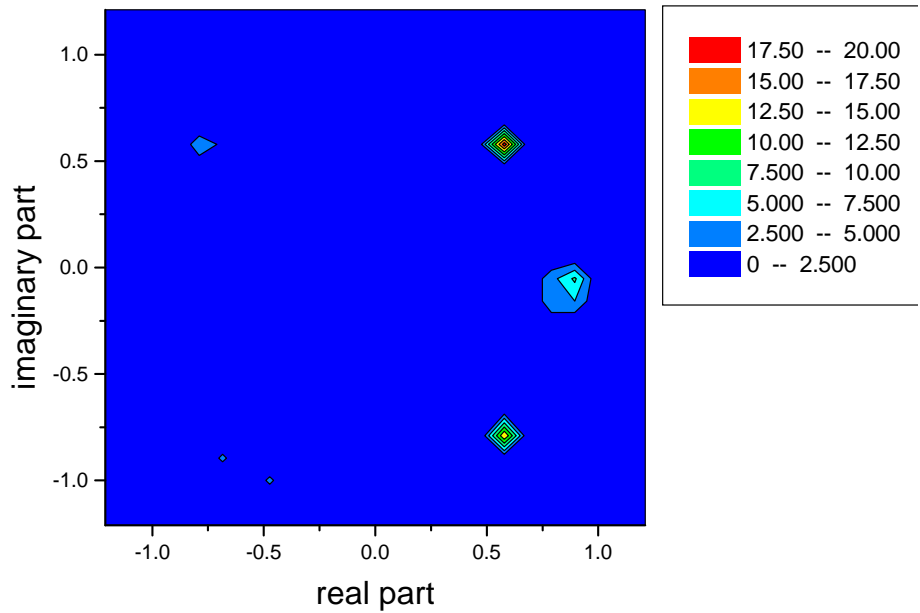,width=0.7\linewidth}
\vspace{-0.9in}
\caption{Signal/noise ratio of approximately $-15db$ for the constant
term and of approximately $-21db$ for the oscillating one.
Distribution of the [149/150] Pad\'{e} approximant poles distant
from their coupled zeros more than $0.2$. The level curves
represent the number of poles in each box of side $0.1$. The two
highest peaks correspond to the oscillating signal poles at
$z=(0.67344,0.67344)$ and $z=(0.67344,-0.67344))$. The
non-oscillating signal pole at $z=(0.95238,0.0)$ appears to be
lower and more spread out.}
\label{fig6}
\end{figure}

Adding to the $-15db$ case an oscillating term, resulting in a couple of complex conjugate poles in
$(0.67344,0.67344)$ and $(0.67344,-0.67344)$ respectively, and having a average signal/noise ratio of approximately
$-21db$ (matched filtering SNR equal to $1.5$), gives us Figure (\ref{fig6}). The oscillating signal, even if
weaker, appears to give peaks much sharper than the peak given by the non-oscillating signal, but only because its
poles are far from box borders: if we sum the four peaks corresponding to the non-oscillating signal we get $19$,
the same as the highest oscillating signal peak (the other one is only $15$, the asymmetry being due to the noise
being complex). Note that the height of the highest ``fake" peak is now $4$, which suggests the need for a higher
false alarm threshold.

\section{ General Conclusion.}

There is still plenty of room for improvement: of the methods outlined in section \ref{s4} we have used, in the
examples given above, only the cleaning of Froissart doublets and the knowledge that the signal poles must be
inside the unit circle.

We are moreover still not using all the information available. For example we did not use the residues of the
poles, from which the signal amplitude can be obtained, and we haven't yet performed a complete optimization over
the various parameters used in the analysis. In particular there is the need for an extensive mapping of the spread
of the reconstructed poles as a function of noise level and Pad\'{e} approximant order. Our last example also
suggests that there might be a connection between number of signals and fake alarm threshold, which needs to be
investigated too.

This said, we can nonetheless conclude the following.

In our method, the Z-transform appears as an extension of the discrete Fourier transform to complex values of the
frequency: in the complex plane the discrete Fourier transform is the restriction of the Z transform to values of z
located at the roots of unity. However, the noise is not uniformly distributed in the complex z-plane and,
unfortunately, the roots of unity are noise attractors as shown at the beginning of the article. Therefore, the
Fourier transform is not a good choice when the signal is embedded in high noise as it is the case in many
circumstances. The analytic treatment of the noise we propose, distinguishes, in a drastic way, the signal from the
noise by their totally different analytic properties. The fact that the noise presents an analytic characterization
that is universal, independent of its statistical properties, (the most novel result of our paper) has as
consequence a much greater independence of the signal identification on the noise level. In principle, in the limit
of very long signals, the signal extraction becomes independent of the signal to noise ratio.

The examples we have given, moreover show that the $Z-transform$ analysis promises {\it not only in theory} to be
an extremely powerful tool especially for {\it the frequently encountered case} of {\bf damped signals in a highly
noisy environment}, as it gives us a sure signature for the signal by radially separating it in the complex plane
from the noise.

\section{Acknowledgments}

We thank Professor Marcel Froissart, from College de France, for discussions and suggestions.

We thank Professor Carlos Handy, Head of the Physics Department at
Texas Southern University, for his support.

We thank Professor Bernhard Beckermann of Lille university, France for explaining to us his method of computing
Pad\'{e} Approximations avoiding divisions.

Special thanks to Professor Mario Diaz, Director of the Center for
Gravitational Waves at the University of Texas at Brownsville:
without his constant support this work would have never been
possible.

\bigskip

Supported by sub-award CREST to the center for Gravitational Waves, Texas University at Brownsville, Texas USA.



\newpage
.

\vspace{0.6in}

\vspace{0.9in}

\newpage
.

\vspace{0.2in}

\vspace{0.6in}

\newpage
.

\vspace{0.2in}

\vspace{0.8in}

\newpage
.

\vspace{0.2in}

\vspace{0.6in}

\end{document}